\documentclass{iaus}
\usepackage{graphicx}
\usepackage{sidecap}
\graphicspath{{./fig/}{./png/}}

\newcommand{\EQ}{\begin{equation}}
\newcommand{\EN}{\end{equation}}
\newcommand{\EQA}{\begin{eqnarray}}
\newcommand{\ENA}{\end{eqnarray}}

\newcommand{\Fig}[1]{Figure~\ref{#1}}

\newcommand{\bra}[1]{\langle #1\rangle}

{}
{}
{}

{}
{}
{}
{}
{}
{}
{}
{}
{}
{}
{}
{}
{}
{}
{}
{}
{}
{}

{}
{}
{}

\newcommand{\meanB}{\overline{B}}

{}

{}
{}

\newcommand{\Rsun}{R}
\newcommand{\gggg}{\mbox{\boldmath $g$} {}}

\newcommand{\rr}{\mbox{\boldmath $r$} {}}

\newcommand{\uu}{\mbox{\boldmath $u$} {}}
\newcommand{\UU}{\mbox{\boldmath $U$} {}}

\newcommand{\BB}{\mbox{\boldmath $B$} {}}

\newcommand{\JJ}{\mbox{\boldmath $J$} {}}

\newcommand{\AAA}{\mbox{\boldmath $A$} {}}

\newcommand{\nab}{\mbox{\boldmath $\nabla$} {}}
\newcommand{\OO}{\mbox{\boldmath $\Omega$} {}}
\newcommand{\oo}{\mbox{\boldmath $\omega$} {}}

\newcommand{\SSSS}{\mbox{\boldmath ${\sf S}$} {}}

\newcommand{\DD}{{\rm D} {}}

\def\Co{\mbox{\rm Co}}

\def\Rm{\mbox{\rm Re}_M}

\def\Rey{\mbox{\rm Re}}

\def\Co{\mbox{\rm Co}}

\def\kf{k_{\rm f}}

\def\urms{u_{\rm rms}}

\def\Beq{B_{\rm eq}}

\def\half{{\textstyle{1\over2}}}

\def\onethird{{\textstyle{1\over3}}}

\newcommand{\yapjl}[3]{ #1, {ApJ,} {#2}, #3}

\title[Coronal ejection driven by convective dynamo action] %% give here short title %%
{Coronal ejections from convective spherical shell dynamos}

\author[Warnecke et al.]   %% give here short author list %%
{J.\ Warnecke$^{1,2}$
 P.\ J.\ K{\"a}pyl{\"a}$^{1,3}$
 M.\ J.\ Mantere$^{3}$
 \and A.\ Brandenburg$^{1,2}$}
\affiliation{$^1$Nordita, Roslagstullsbacken 23,
SE-10691 Stockholm, Sweden, email: {\tt joern@nordita.org} \\
$^2$Department of Astronomy, Stockholm University,
SE 10691 Stockholm, Sweden\\
$^{3}$ Department of Physics, PO BOX 64, FI-00014 Helsinki University, Finland}

\pubyear{2012}
\volume{286}  %% insert here IAU Symposium No.
\pagerange{1--5}
\jname{Comparative magnetic minima}
\editors{C. H. Mandrini, eds.}
\doi{}

\begin{document}

\maketitle

\begin{abstract}
We present a three-dimensional model of rotating convection combined with a
simplified corona in spherical coordinates.
The motions in the convection zone generate a large-scale magnetic field
that gets sporadically ejected into the outer layers above.
Our model corona is approximately isothermal, but it includes density
stratification due to gravity.

\end{abstract}
\firstsection % if your document starts with a section,
              % remove some space above using this command.
\section{Introduction}
The Sun sheds plasma into the heliosphere via coronal mass
ejections (CMEs).
There has been significant progress in the study of CMEs in recent years.
In addition to improved observations from spacecrafts like SDO
or STEREO, there have also been major advances in the field of
numerical modeling of CME events.
One of the main motivations for understanding the generation and
dynamics of CMEs is to have more reliable predictions for space
weather. CMEs can have strong impacts on Earth and can affect
microelectronics aboard spacecrafts.
However, an important side effect of CMEs is that the Sun sheds
magnetic helicity from the convection zone which may prevent the
solar dynamo from being quenched at high magnetic Reynolds numbers (Blackman \&
Brandenburg, 2003).

In most of the CME models, the motion in the photosphere as well as the initial
and footpoint magnetic fields are prescribed or taken from two-dimensional
observations at the solar surface.
Such fields represent an incomplete approximation to the full
three-dimensional field generated by the dynamo.
An alternative way of modeling CMEs would be to perform a 3-D
convection simulation to generate the configuration of magnetic field
and the
photospheric motions self-consistently.
However, the convection zone and the solar corona have very different
timescales.
In solar convection the dominant timescale varies from minutes to days.
While this is short compared with the dynamo cycle, time scales in the
solar corona can be even shorter because the Alfv\'en speed is large.

In earlier work (Warnecke \& Brandenburg 2010, Warnecke et
al.\ 2011) we have established a two-layer model with a unified treatment of the
convection zone and the solar corona in a single three-dimensional domain.
In those models, the generation of magnetic fields was driven by
turbulence from random forcing with helical
transverse waves mimicking the effects of convection and rotation in a
simplified way.
This two-layer model was able to produce recurrent
plasmoid ejections which are similar to observed eruptive features on the Sun.
In this work we develop this approach further and apply a
self-consistent convection model instead of random forcing.
This model automatically includes differential rotation as a result of the
interaction of rotation and convection.
Here we present some preliminary results from such a study.
We find the formation of a large-scale magnetic field, which 
eventually gets ejected as a magnetic
structure.
It seems to us that this mechanism could play an important role for the
formation of CMEs and flare events on the Sun and other stars. 
\section{The model}
\label{model}
As in Warnecke \& Brandenburg (2010) and Warnecke et al.\ (2011) a
two-layer model is used.
Our convection zone is similar to the one in K\"apyl\"a
et al.\ (2010, 2011).
The domain is a segment of the Sun and is described in spherical polar
coordinates $(r,\theta,\phi)$.
We mimic the convection zone starting at radius $r=0.7\,\Rsun$ and
the solar corona until $r=1.5\,\Rsun$,
where $\Rsun$ denotes the solar radius, used from here on
as our unit length.
In the latitudinal direction, our domain extends from colatitude
$\theta=15^{\circ}$ to $165^{\circ}$ and in the azimuthal direction
from $\phi=0^{\circ}$ to $90^{\circ}$.  We solve the
equations of compressible magnetohydrodynamics,
\begin{eqnarray}
{\partial\AAA\over\partial t}&=&\UU\times\BB+\eta\nabla^2\AAA,\\
{\DD\ln\rho\over \DD t} &=&-\nab\cdot\UU,\\
{\DD\UU\over\DD t}&=&  \gggg - 2\OO \times \UU + {1\over\rho}
\left(\JJ\times\BB - \nab p+\nab\cdot 2\nu\rho\SSSS\right),\\
T{\DD s\over\DD t}&=&{1\over\rho}\nab\cdot K\nab T +
2\nu\SSSS^2+{\mu_0\eta\over\rho}\JJ^2 - \Gamma_{\rm cool},
\end{eqnarray}
where the magnetic field is given by $\BB=\nab\times\AAA$ and thus obeys
$\nab\cdot\BB=0$ at all times.
The vacuum permeability is given by $\mu_0$, whereas magnetic
diffusivity and kinematic viscosity are given by $\eta$ and $\nu$,
respectively.
${\mathsf S}_{ij}=\half(U_{i;j}+U_{j;i})-\onethird\delta_{ij}\nab\cdot\UU$
is the traceless rate-of-strain tensor, and semicolons denote covariant
differentiation,
$\OO =\Omega_0(\cos\theta,-\sin\theta,0)$ is the rotation vector,
$K$ is the radiative heat conductivity and $\gggg=-GM\rr/r^3$ is the
gravitational acceleration.
The fluid obeys the ideal gas law with $\gamma=5/3$.
We consider a setup in which the stratification is convectively
unstable below $r=Rsun$, whereas the region above is stably
stratified and isothermal due to a cooling term $\Gamma_{\rm cool}$ in
the entropy equation.  The $\Gamma_{\rm cool}$ term is $r$ dependent and
causes a smooth transition to the isothermal layer representing the
corona.

The simulation domain is periodic in the azimuthal direction. 
For the velocity we use
stress-free conditions at all other boundaries.
For the magnetic field we adopt radial field conditions on the
$r=1.5\,\Rsun$ boundary and perfect conductor conditions on the $r=0.7\,\Rsun$ and
both latitudinal boundaries.
Time is expressed in units of $\tau = \left(\urms\kf\right)^{-1}$, which is
the eddy turnover time in the convection zone. 
We use the 
{\sc Pencil Code}\footnote{\texttt{http://pencil-code.googlecode.com}},
which uses sixth-order centered finite differences in space and 
a third-order Runge-Kutta scheme in time;
see Mitra et al.\ (2009) for the extension to spherical coordinates. 
\section{Results}
\begin{figure}[t!]
\begin{center}
\includegraphics[width=0.49\columnwidth]{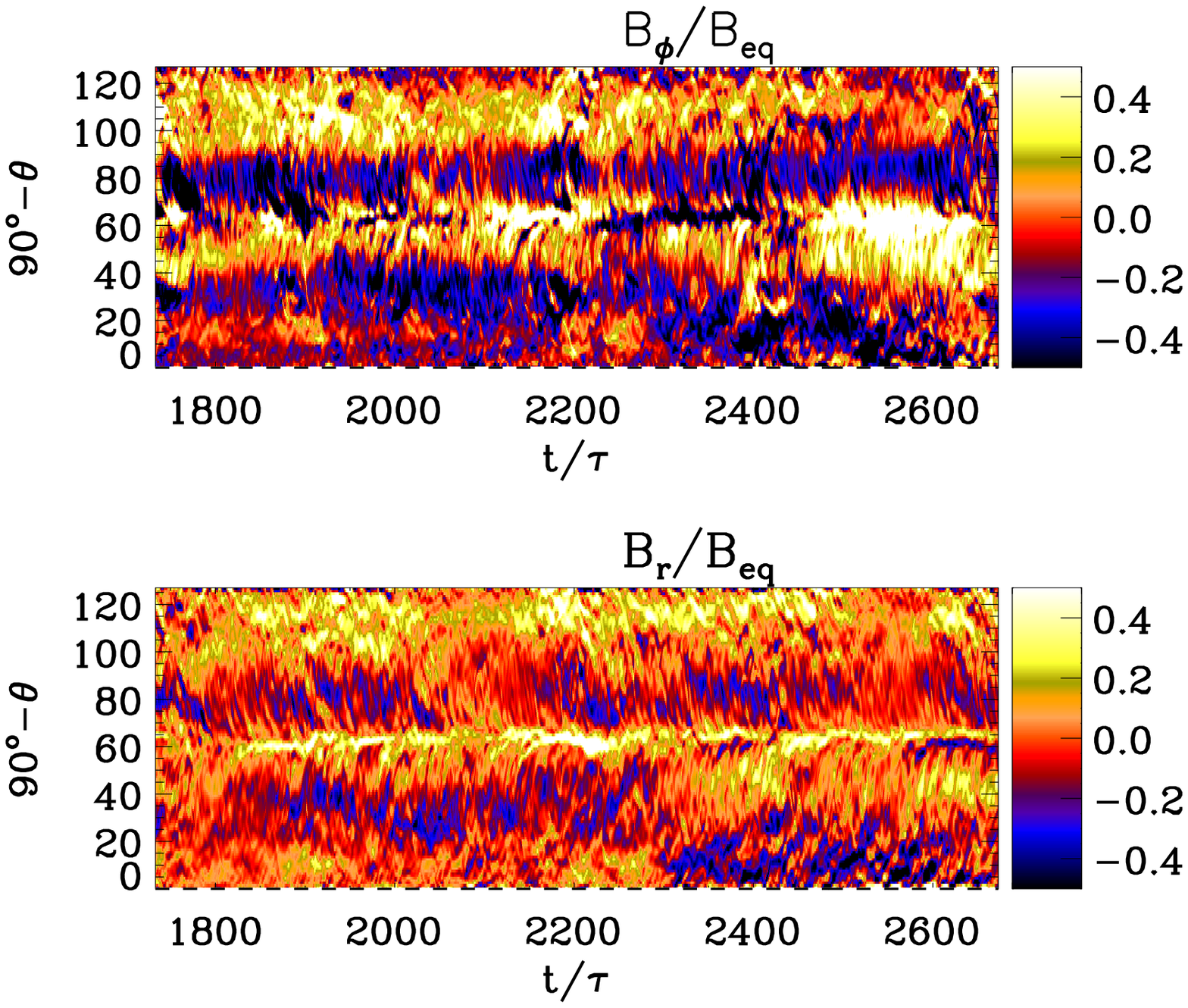}
\includegraphics[width=0.49\columnwidth]{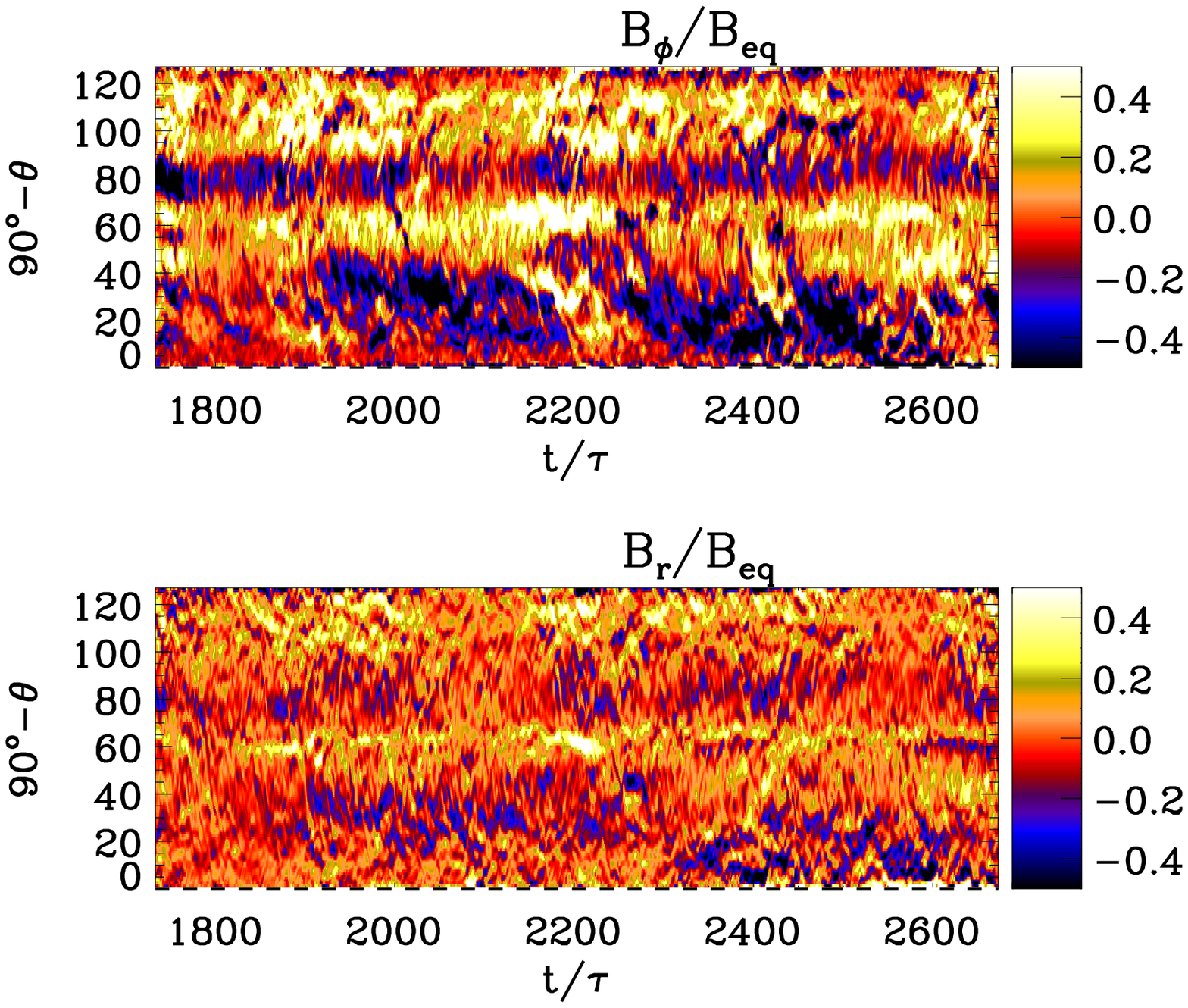}
\end{center}\caption[]{
Variation of $\meanB_{\phi}$ and $\meanB_{r}$ 
in the convection zone at $r=0.89\Rsun$ (left panel) and
$r=0.79\Rsun$ (right).
Dark blue shades represent negative and light yellow positive values.
The dotted horizontal lines show the location of the equator at
$\theta=\pi/2$.
The magnetic field is normalized by the equipartition value.
}
\label{but1}
\end{figure}

\begin{figure}[t!]
\begin{center}
\includegraphics[width=0.49\columnwidth]{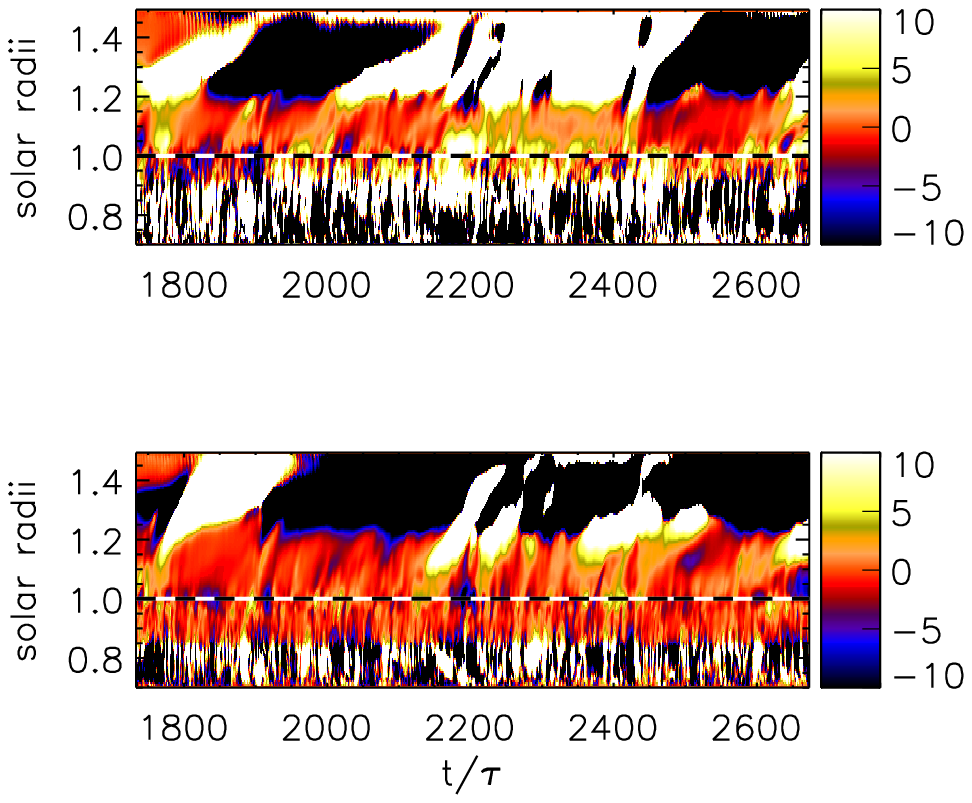}
\includegraphics[width=0.49\columnwidth]{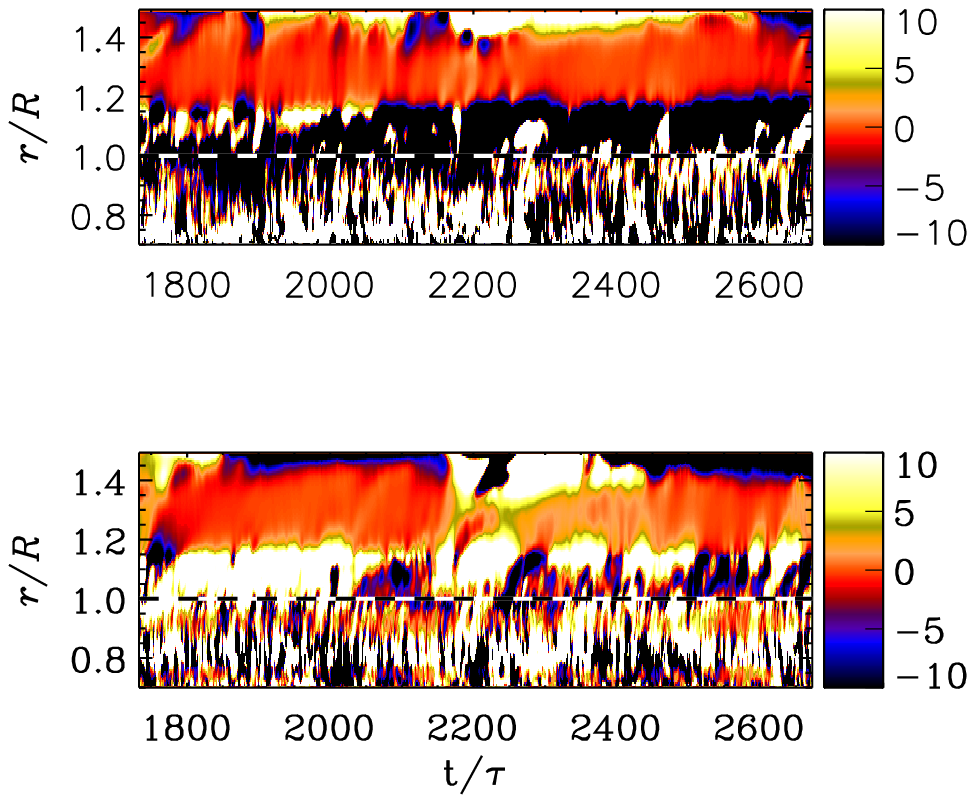}
\end{center}\caption[]{
Dependence of the dimensionless ratio
$\mu_0 R\,\overline{\JJ\cdot\BB} / \bra{\overline{\BB^2}}_t$
on time $t/\tau$ and radius $r$ in terms of the solar radius. 
The top panels show a narrow band in $\theta$ in the northern
hemisphere and the bottom ones in the southern hemisphere.
We have also averaged in latitude from
$4.1^{\circ}$ to $19.5^{\circ}$ (left panel) and $32.5^{\circ}$ to
$45.5^{\circ}$ (right).
Dark blue shades represent negative and light yellow positive values.
The dotted horizontal lines show the location of the surface at $r=\Rsun$.
}
\label{pjb}
\end{figure}

In this work we focus on a run which has
fluid Reynolds number $\Rey=3$, magnetic Reynolds number
$\Rm=32$ and Coriolis number $\Co=7$.
We define the Reynolds number as $\Rey=\urms/\nu\kf$ for the magnetic
one $\Rm=\urms/\eta\kf$, respectively and the Coriolis number as $\Co=2\OO/\urms\kf$.
After around 100 turnover times, the onset of large-scale dynamo
action due to the convective motions is observed. 
The magnetic field
reacts back on the fluid motions and causes saturation after around 200
turnover times.  The saturation is combined with an oscillation of the
magnetic field strength in the convection zone.  
The field reaches
its maximum strength of about 60\% of the equipartition field
strength, $(\Beq=(\mu_0\overline{\rho\uu^2})^{1/2}$, which is
comparable with the values obtained in the forced turbulence counterparts
both in Cartesian and spherical coordinates (Warnecke \& Brandenburg
2010, Warnecke et al.\ 2011). 
The magnetic field in rotating
convection seems to show certain migration properties.  In \Fig{but1},
we show the azimuthal ($\meanB_{\phi}$) and radial ($\meanB_{r}$)
magnetic fields versus time ($t/\tau$) and latitude
($90^{\circ}-\theta$) for two different heights.
\begin{figure*}[t!]
\begin{center}
\includegraphics[width=2.6cm]{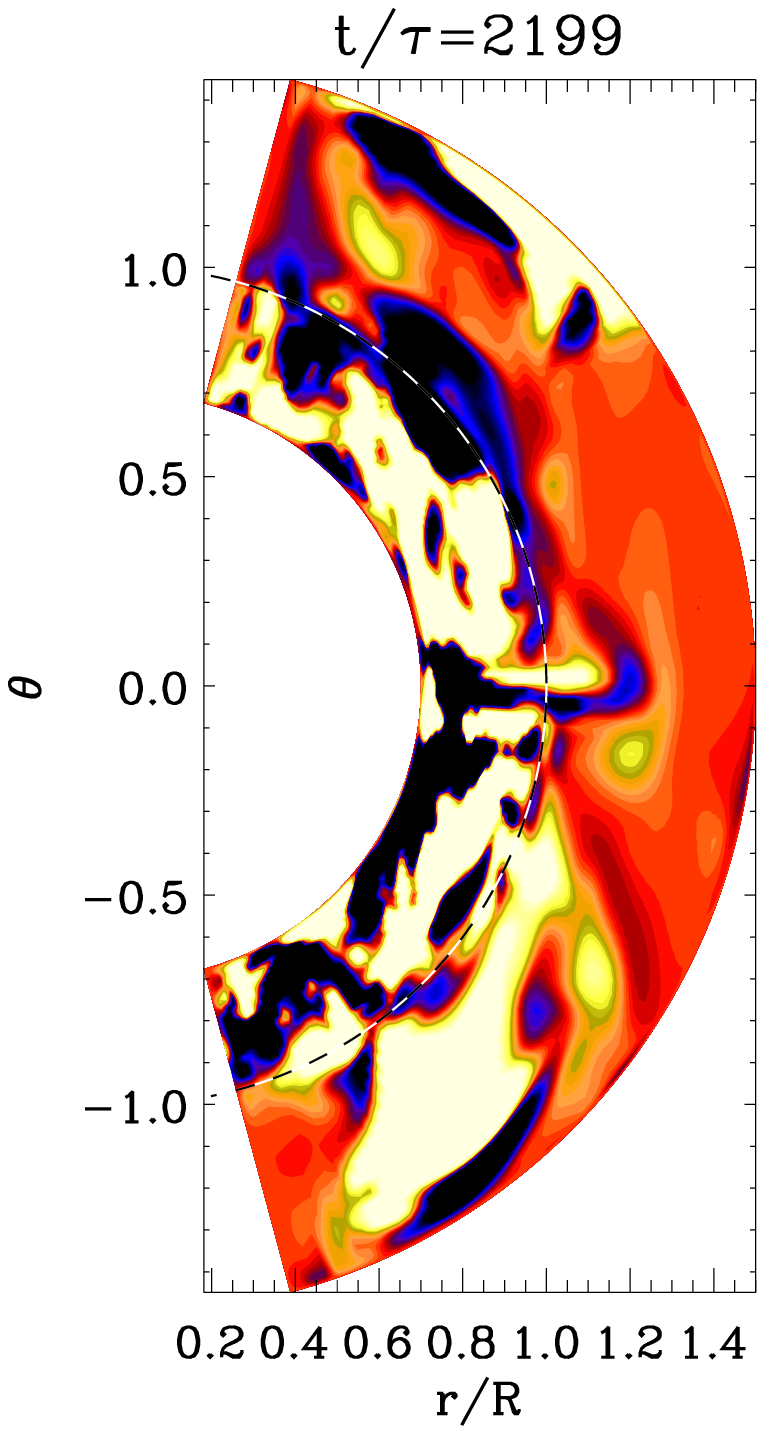}
\includegraphics[width=2.6cm]{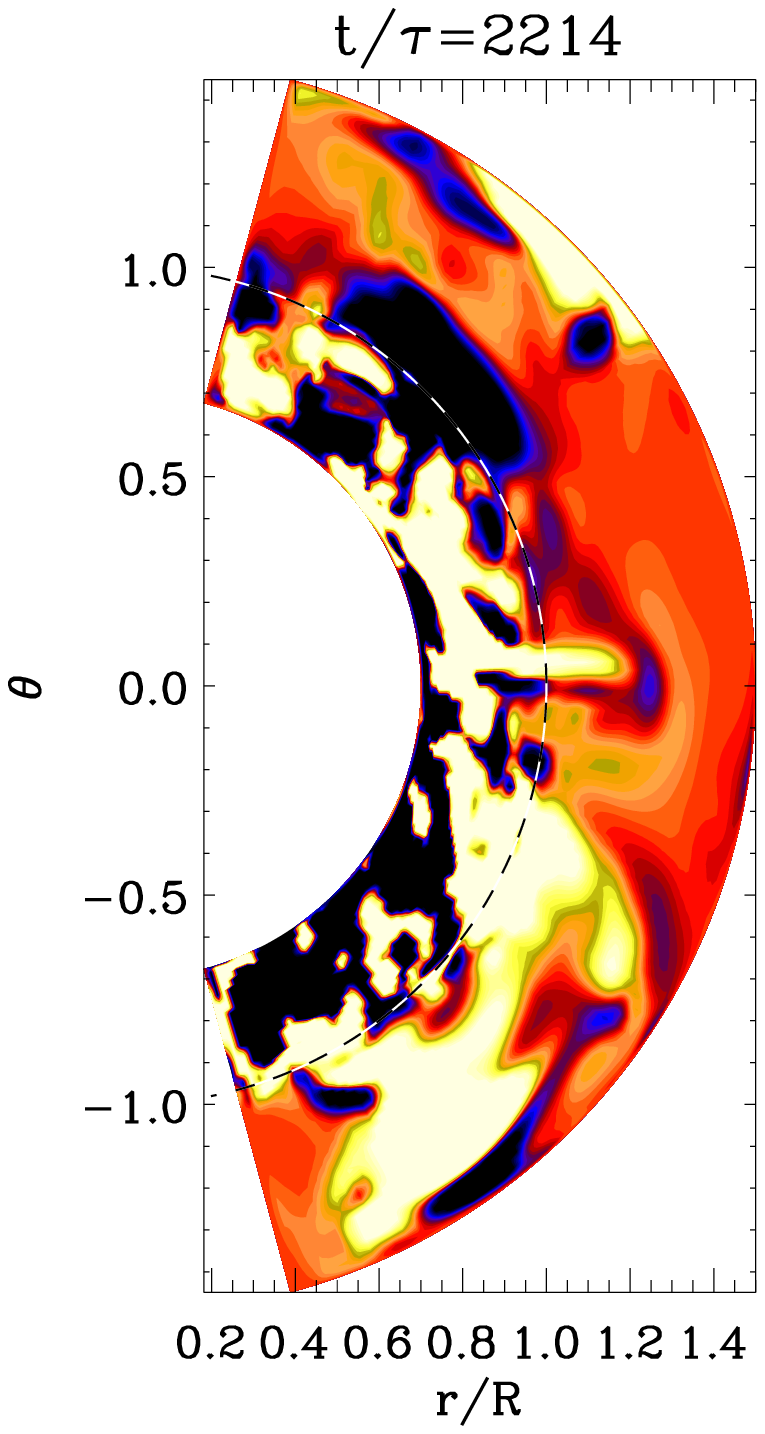}
\includegraphics[width=2.6cm]{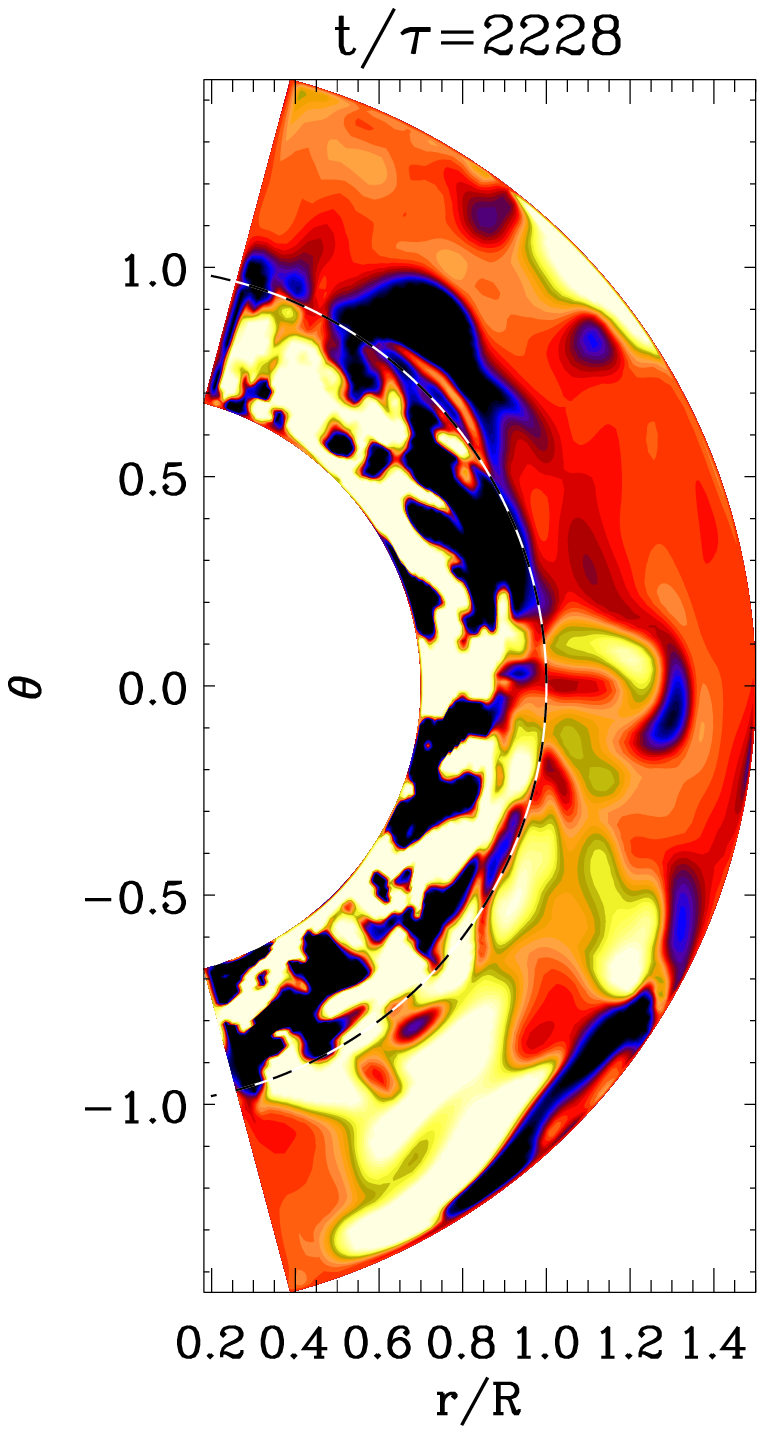}
\includegraphics[width=2.6cm]{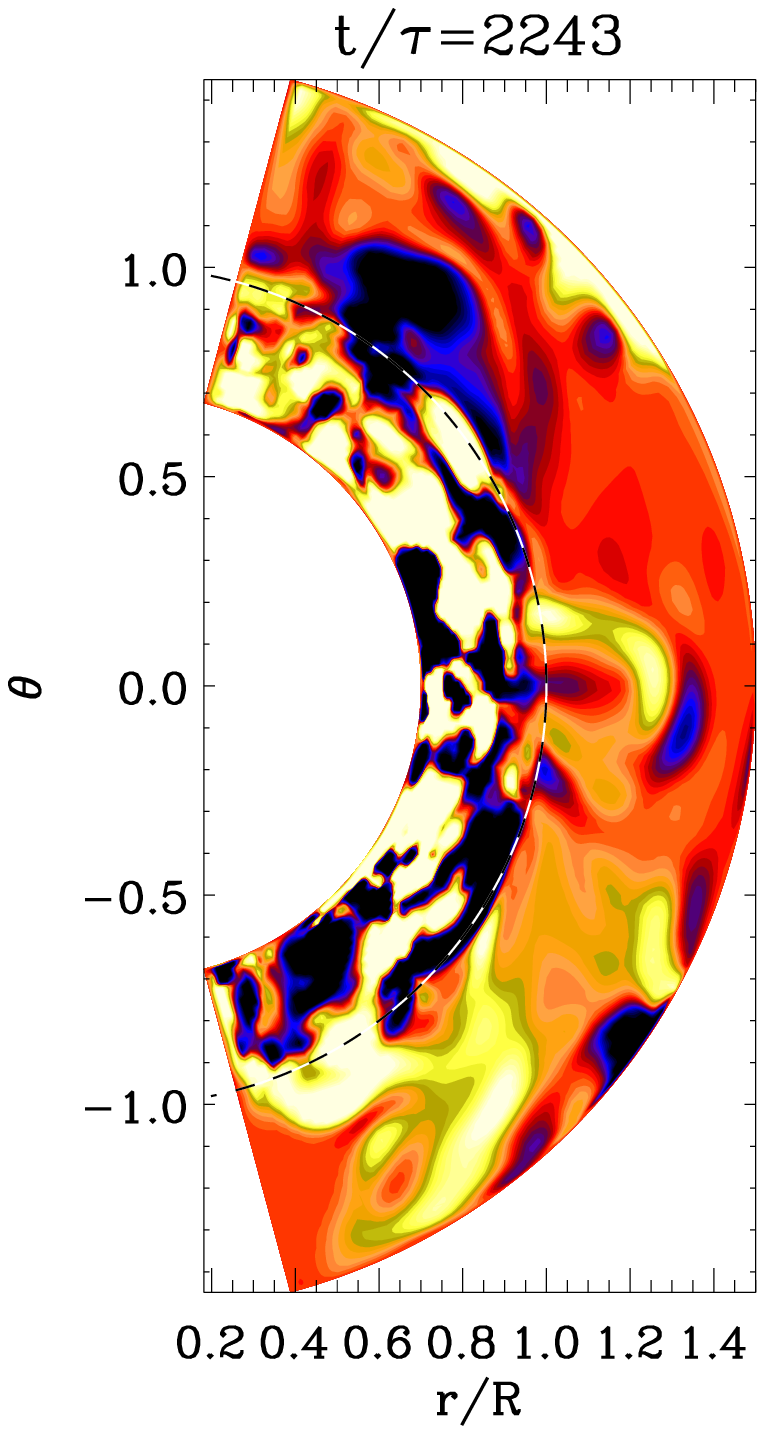}
\includegraphics[width=2.6cm]{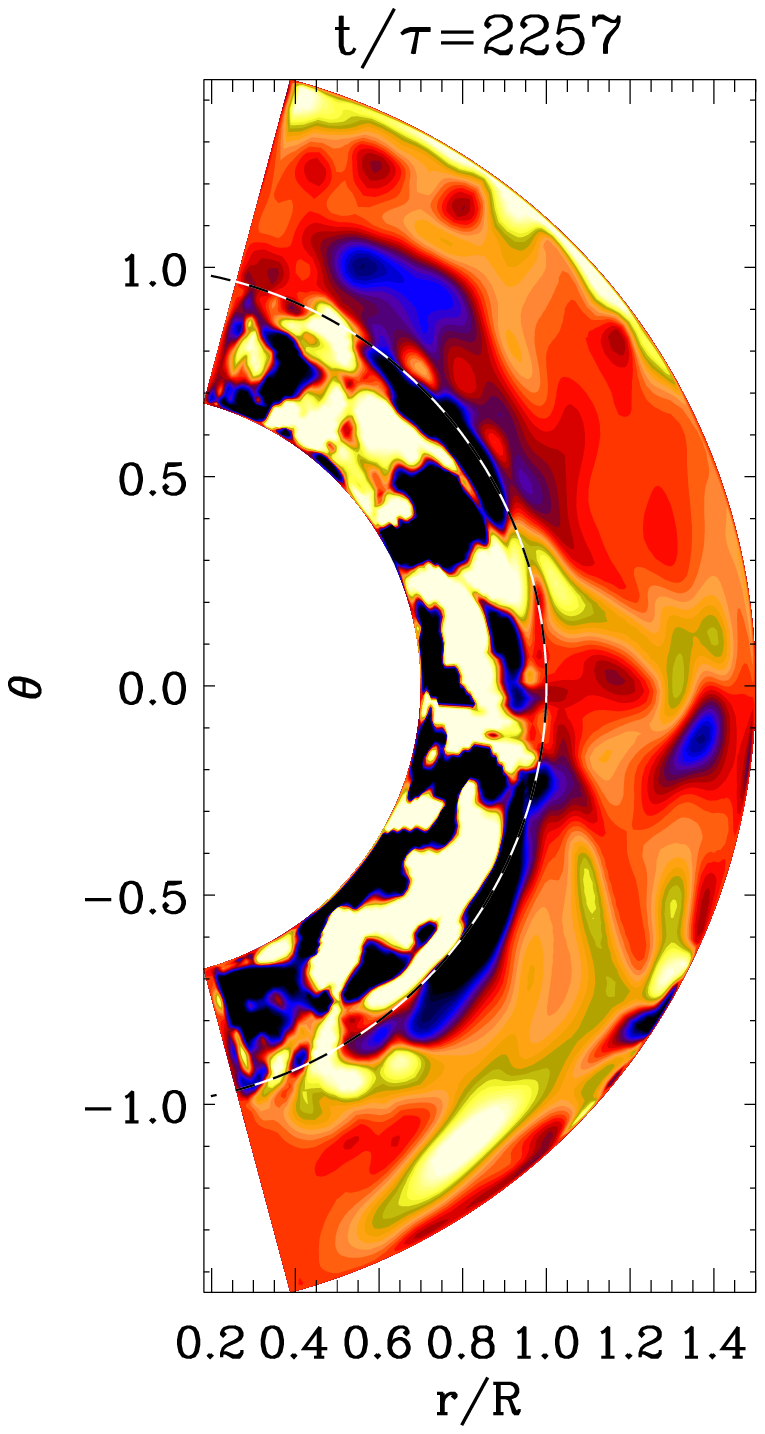}
\end{center}\caption[]{
Time series of coronal ejections in spherical coordinates.
The normalized current helicity,
$\mu_0 R\,\overline{\JJ\cdot\BB}/\bra{\overline{\BB^2}}_t$, is shown in a
color-scale representation from different times; dark blue represents
negative and light yellow positive values.
The dashed horizontal lines show the location of the surface at $r=\Rsun$.
}
\label{jb}
\end{figure*}
The magnetic field emerges through the surface and is ejected as
isolated structures.
The dynamical evolution is clearly seen in the sequence of images of
\Fig{jb}, where the normalized current helicity ($\mu_0
R\,\overline{\JJ\cdot\BB}/\bra{\overline{\BB^2}}_t$) is shown.
If one focuses on the region near the equator ($\theta=\pi/2$), a small
yellow (i.e.\ positive) feature with a blue (negative) arch emerges
through the surface to the outer atmosphere, where it leaves the
domain through the outer boundary.
This ejection does not occur as a single event---it rather shows recurrent
behavior.
We do not, however, find a clear periodicity in the ejection recurrence, like in
earlier work.
Even though the ejected structures are much smaller than in Warnecke et
al.\ (2011), their shape is similar.
However, the detection of an ejection with the aid of the current helicity
is much more difficult in convection-driven simulations than in forced
turbulence.
In \Fig{jb}, one can see large structures diffusing through the surface
into the upper atmosphere at higher latitudes.
These structures disturb the emergence of ejections
and hamper their detection.
These larger diffusive structures are also visible in \Fig{pjb}, where
the normalized current density is averaged over two narrow latitude bands
on each hemisphere.
The formation of these diffuse structures in the corona seem to get
suppressed when the stratification of the system is
increased. Nevertheless, ejections are still visible, for example
around $t/\tau=1900$, $t/\tau=2200$ (see \Fig{jb}) and $t/\tau=2400$.

The hemispheric rule, as we have found in earlier work,
is not completely valid in the current work.
The current helicity in \Fig{pjb} shows a different behavior in the
northern and southern hemispheres, but one cannot tell clearly the
leading sign.
This has to do with the much lower values of relative kinetic helicity
$h_{\rm rel}(r,t)=\bra{\oo\cdot\uu}/\omega_{\rm rms}\urms$
in the convection simulations.
Values of up to $h_{\rm rel}=\mp0.4$ are reached at some radii
in the two hemispheres.
In the forced turbulence simulations of our earlier work, we studied
purely helical systems with nearly $h_{\rm rel}=\mp1$.

In summary, we are able to advance our two-layer model approach by
including self-consistent convection, which generates the magnetic field
and eventually drives ejections.
Not surprisingly, the ejections occur non-periodically and as smaller
structures than in earlier work, but this behavior seems to be similar to
the Sun.
Furthermore, detailed investigations covering a wider range of
magnetic and kinetic Reynolds number as well as rotation rates show
promising results.

\end{document}